\documentstyle[prb,aps,floats,graphicx]{revtex}
\begin{document}
\pagestyle{empty}

\twocolumn[\hsize\textwidth\columnwidth\hsize\csname
@twocolumnfalse\endcsname

\title{Frustrated 3-Dimensional Quantum Spin Liquid in CuHpCl}

\author{M. B. Stone$^{1}$, Y. Chen$^{1}$, J. Rittner$^{1}$, H.
Yardimci$^{1}$, D. H. Reich$^{1}$, C.
Broholm$^{1,2}$, D. V. Ferraris$^{3}$, and T. Lectka$^{3}$}

\address{
$^{1}$Department of Physics and Astronomy, The Johns Hopkins
University,
Baltimore, MD 21218
\\
$^{2}$National Institute of Standards and Technology, Gaithersburg, MD
20899
\\
$^{3}$Department of Chemistry, The Johns Hopkins University,
Baltimore, MD
21218}

\date{\today}
\maketitle
\begin{abstract}
Inelastic neutron scattering measurements are reported for
the quantum antiferromagnetic material
$\rm Cu_2(C_5H_{12}N_2)_2Cl_4$ (CuHpCl).
The magnetic excitation spectrum forms a band extending from 0.9 meV
to 1.4 meV.  The spectrum contains two modes that disperse throughout
the $\bf a-c$ plane of the monoclinic unit cell with less dispersion
along the unique $\bf b$-axis. Simple arguments based on the measured
dispersion relations and the crystal structure show that a spin
ladder model is inappropriate for describing CuHpCl. Instead, it is
proposed that hydrogen bond mediated exchange interactions between
the bi-nuclear molecular units yield a three-dimensional interacting
spin system with a recurrent triangular motif similar to the
Shastry-Sutherland Model (SSM). Model-independent analysis based on
the first moment sum rule shows that at least four distinct spin
pairs are strongly correlated and that two of these, including the
dimer bond of the corresponding SSM, are magnetically frustrated.
These results show that CuHpCl should be classified as a frustration
induced three dimensional quantum spin liquid.

\end{abstract}
\narrowtext
\vskip0pc]
\newpage

\section{Introduction}

Organometallic magnets\cite{miller} are excellent model systems in
which to explore the intricate quantum many-body physics of
interacting spin systems.\cite{natodyn}
They are attractive because their energy scales are well matched to
efficient experimental probes of magnetism and because a wide range
of magnetic phases are found in these materials. In addition to
supporting frozen magnetic states with ferromagnetism,
ferrimagnetism,\cite{asano} antiferromagnetism and
meta-magnetism,\cite{miller} organometallic  magnets also provide
intriguing examples of quantum spin
liquids.\cite{organqsl,geilopaper}  These can be defined as strongly
correlated states of interacting spin systems where time reversal
symmetry persists at temperatures far below the characteristic energy
scale for interactions.

Quantum spin liquids are most commonly found in quasi-one-dimensional
antiferromagnetic systems such as the uniform spin-1
chain,\cite{nenp,ndmap} the alternating spin
chain,\cite{Xu00,Johnston87,Eccleston94,Garrettprb97,Garrettprl97}
and the spin-ladder.\cite{Carter96,Eccleston98,Matsuda99,watson2001}
Low dimensionality generally enhances the phase space for low energy
fluctuations, \cite{merminwagner} and in one dimension,
the effect is to preclude a
frozen state involving Heisenberg spins. \cite{geilopaper}  However,
there are also examples of spin-dimer systems where the singlet
ground state associated with antiferromagnetically interacting spin
pairs survives the effects of weaker inter-dimer interactions in two
and three
dimensions.\cite{Leuenberger84,Sagaso97,Cavadini99,Cavadini00}

Geometrical frustration is an alternate route to strong
fluctuations\cite{geilopaper,Ramirez} and there are theoretical
predictions\cite{anderson} of quantum\cite{cana98,palmer} and
classical\cite{moes98} spin liquids based on this effect. While
materials that approximate kagom\'{e}\cite{scgo,qsferrite} and
pyrochlore\cite{zncro,ymoo} antiferromagnets generally have a
freezing transition at sufficiently low temperatures, there are other
more complex structures where geometrical frustration stabilizes an
isolated singlet ground state in the $T=0$
limit.\cite{taniguchi95,Kageyama99,stone00} $\rm SrCu_2(BO_3)_2$ is a
three-dimensional version of the so-called Shastry-Sutherland
model\cite{ssmodel} where the inter-dimer interactions are almost as
strong as the intra-dimer interactions, and yet they fail to induce
gapless magnetic fluctuations and long range order because no such
state can satisfy all interactions.\cite{Kageyama99,kageyama00}
Frustration is also central to stabilizing the quasi-two-dimensional
cooperative singlet state in the organometallic  magnet
PHCC.\cite{stone00} While in $\rm SrCu_2(BO_3)_2$ correlations exist
only within spin-dimers,\cite{kageyama00} the correlated spin
clusters in PHCC involve at least 8 spins of which two spin pairs are
frustrated and provide a positive contribution to the ground state
energy. PHCC was previously identified as an alternating spin chain
based on susceptibility measurements. This and other
misidentifications\cite{Garrettprb97,Garrettprl97} indicate that
conventional bulk measurements cannot reliably determine the origin
and nature of spin systems with a gapped excitation spectrum and that
more sophisticated techniques should be applied to explore these
unique systems.

\begin{figure}
    \centering\rotatebox{270}{\includegraphics[scale=0.30]{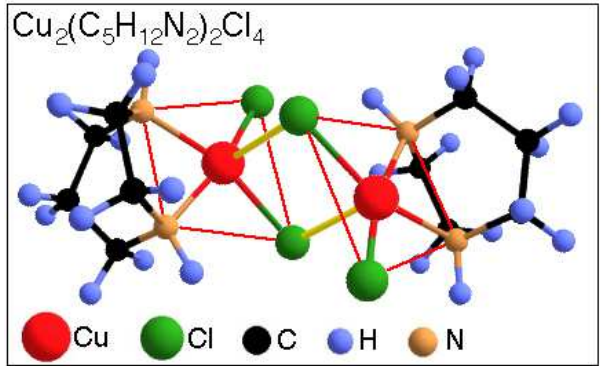}}
\caption{\label{fig:onemolecule}
The molecular formula unit of CuHpCl featuring two copper atoms in
4+1 square pyramidal coordination.\protect\cite{Chiari90}
The apical Cu-Cl bonds are shown in yellow.
The Cu coordination pyramids share an apical edge
and have parallel basal planes (red lines).}
\end{figure}

One magnet with a gapped excitation spectrum
that has received considerable attention is
$\rm Cu_2(C_5H_{12}N_2)_2Cl_4$
(Cu$_{2}$(1,4-diazacycloheptane)$_{2}$Cl$_{4}$  , or for short
CuHpCl).
\cite{Chiari90,Hammar96,ChabouPRB97,ChabouPRL97,ChabouPRL98,Hammar98,ChabouEJPB98,Chiba98,Hagiwara98,Deguchi98,Calemczuk99,Ohta99,ChabPhysicaB00,Mayaffre00}
Measurements of the magnetization, magnetic susceptibility,
and specific heat show that this system has
a spin gap $\Delta \approx 0.9 $ meV,
a magnetic bandwidth of approximately 0.5 meV,
and a saturation field $H_{c2} = 13.2$ T.
Based on these measurements and the crystal
structure, it was proposed that CuHpCl is a two-leg
spin ladder composed of coupled dimers, with the
intra-dimer bonds of strength  $J_{1} = 1.14$ meV
forming the rungs, and inter-dimer bonds of strength
$J_{2} \approx 0.2 J_{1}$ forming the legs of the ladders.
\cite{ChabouPRB97}
Subsequent experimental
results have generally been interpreted in terms of this model,
and the system has been the inspiration for a number
of theoretical studies.
\cite{Elstner98,Giamarchi99,Wang00,Normand00,Hayward96,Usami98,Gu99,Gu99b,Wei97}

Despite this extensive body of work, the true nature of the spin
interactions in CuHpCl has never been conclusively
demonstrated.  Susceptibility and
specific heat data are equally well described by spin-ladder,
alternating chain, and coupled-bilayer models.  \cite{Elstner98}
Some evidence that the two-leg ladder model might in fact not be
appropriate for CuHpCl came from a previous inelastic neutron
scattering experiment \cite{Hammar98} on a powder sample, where we
found that the wavevector-integrated magnetic scattering intensity
did not show the characteristic van Hove singularities expected for
the magnetic density of states of a one-dimensional spin system.  In
this measurement, the wavevector dependence of the energy-integrated
magnetic scattering was also inconsistent with the predictions of the
ladder model.  However, conclusions from this experiment were limited
by the coarse wavevector resolution of the instrument employed, and
the large,
non-magnetic background signal from the hydrogenous sample.

In this paper, we report inelastic neutron scattering measurements
performed both on a deuterated powder with improved wavevector
resolution and on hydrogenous single crystals.  The powder and
single-crystal measurements are each independently inconsistent with
the spin-ladder model.  The powder data indicate that the strongest
dimer bond is different from that of the ladder model.  The single
crystal measurements show the presence of two modes within the 0.5 meV
bandwidth, a feature not predicted by the spin-ladder model.
Consideration of the measured dispersion relation and the structure of the
system
lead to the conclusion that the network of significant magnetic
interactions in CuHpCl is three dimensional.  Model-independent
analysis of the powder and single-crystal data based on the first moment
sum-rule provide the contributions of each crystallographically
distinct spin pair to the ground state energy.  Two classes of spin
pairs yield positive contributions, indicating that geometrical
frustration plays an important role in stabilizing the quantum spin
liquid in CuHpCl.
\begin{figure}
    \centering\includegraphics[scale=0.85]{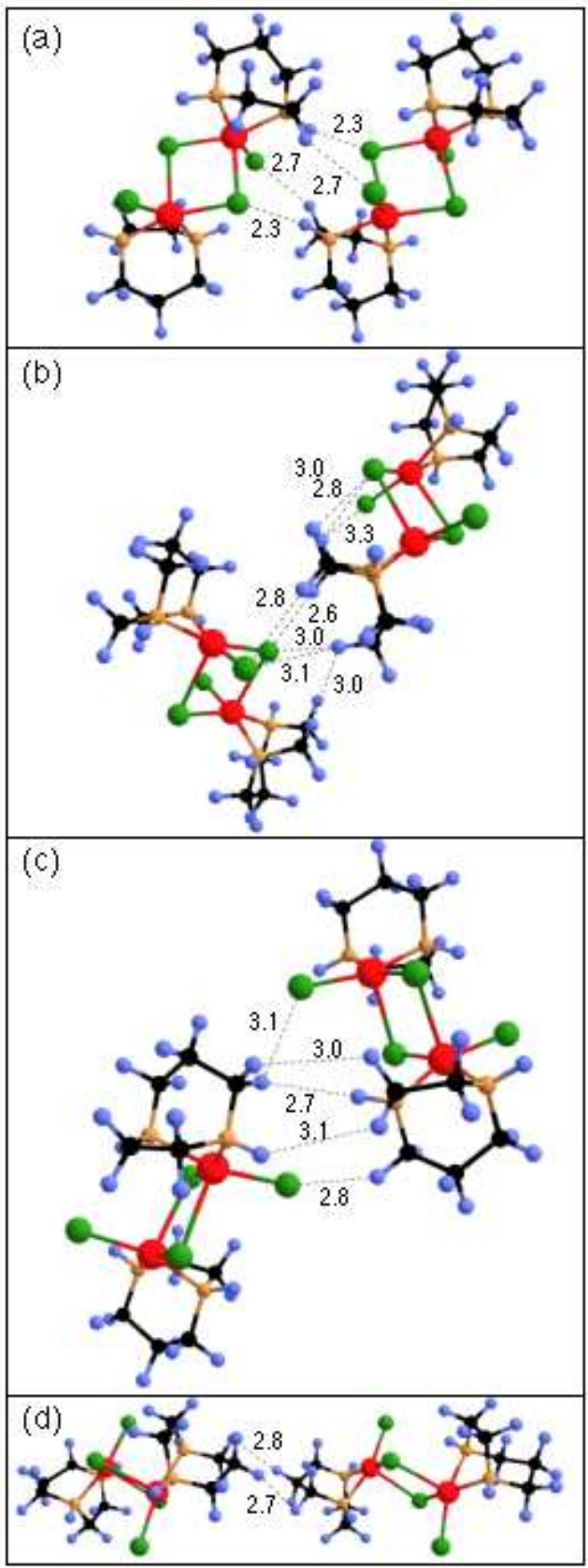}
\caption{\label{fig:twomolecules}
The four different types of near neighbor hydrogen bonded molecular
pairs in CuHpCl.\protect\cite{Chiari90} Numbers indicate hydrogen
bond distances in \AA . The orientations of the molecules correspond
to the lattices of interactions shown in Fig.~\ref{fig:lattices}.
Color coding is the same as for Fig.~\ref{fig:onemolecule}.
}
\end{figure}

\subsection{Structure of CuHpCl}
\label{structure}

CuHpCl is monoclinic with space group $P2_{1}/c$ and room temperature
lattice constants $a = 13.406(3)$ \AA , $b=11.454(2)$ \AA ,
$c=12.605(3)$ \AA, and
$\beta = 115.01(2)^{\circ}$. \cite{Chiari90}  The low-temperature
lattice constants measured with neutron scattering at $T = 4$ K are
$a = 13.35(4)$ \AA , $b= 11.24(6)$ \AA , $c=12.72(4)$ \AA , and
$\beta = 115.2(2)^{\circ}$.
Figure \ref{fig:onemolecule} shows the approximately centro-symmetric
binuclear molecular unit containing the spin pair denoted by bond 1
in Table 1. Each Cu$^{2+}$ ion is in a (4+1) square pyramidal
coordination with a Cl atom at the apex and the base formed by two N
and two Cl atoms. Within the Cu($\mu$-Cl)$_2$Cu complex of CuHpCl,
the coordinating pyramids share apical edges and have parallel basal
planes. Susceptibility measurements for a series of compounds with
this atomic configuration indicate intra-molecular exchange constants
ranging from -1.5 meV (ferromagnetic) to 0.9 meV (AFM) with no
apparent correlation throughout the range of molecular structural
parameters.\cite{Rodriguez}


\begin{table}
\begin{tabular}{cllllcl}
	\multicolumn{1}{c}{Bond ID} &
	\multicolumn{1}{c}{$d$ (\AA)}     &
	\multicolumn{1}{c}{$x/a$}	        &
	\multicolumn{1}{c}{$\pm y/b$}   &
	\multicolumn{1}{c}{$z/c$}         &
	\multicolumn{1}{c}{$J_{\bf d}\langle{\bf S}_{0}\cdot{\bf S}_{\bf
d}\rangle$ (meV)}
	\\ \hline

1   & 3.376 & 0.179 & 0.231 &  0.089 &  0.42(3)   \\
2a  & 5.757 & 0.331 & 0.234 &  0.399 &    \\
2b  & 5.813 & 0.312 & 0.227 &  0.423 & \raisebox{1.9ex}[0pt]{0.06(4)}
\\
3a  & 6.987 & 0.509 & 0.003 &  0.488 &    \\
3b  & 7.000 & 0.491 & 0.003 &  0.512 & \raisebox{1.9ex}[0pt]{-0.29(3)} \\
4a  & 7.024 & 0     & 0.266 &  0.5   &    \\
4b  & 7.057 & 0     & 0.273 &  0.5   & \raisebox{1.9ex}[0pt]{-0.18(1)} \\
5a  & 7.154 & 0.491 & 0.269 &  0.012 &    \\
5b  & 7.502 & 0.509 & 0.269 & -0.012 & \raisebox{1.9ex}[0pt]{0.06(2)}
\\
6a  & 7.303 & 0.312 & 0.5   & -0.077 &    \\
6b  & 7.586 & 0.331 & 0.5   & -0.101 & \raisebox{1.9ex}[0pt]{-0.05(4)} \\
7a  & 8.648 & 0.179 & 0.497 & -0.411 &    \\
7b  & 8.698 & 0.179 & 0.503 & -0.411 & \raisebox{1.9ex}[0pt]{-0.09(7)} \\
8a  & 8.814 & 0.179 & 0.497 &  0.589 &    \\
8b  & 8.863 & 0.179 & 0.503 &  0.589 & \raisebox{1.9ex}[0pt]{-0.15(3)} \\
9   & 8.910 & 0.179 & 0.769 &  0.089 &  0.1(2)   \\
10a & 9.327 & 0.669 & 0.234 &  0.601 &    \\
10b & 9.343 & 0.688 & 0.227 &  0.577 & \raisebox{1.9ex}[0pt]{0.01(6)}
\\
\end{tabular}

\caption{Cu-Cu Bond lengths and fractional coordinates for CuHpCl
calculated from previously determined atomic coordinates
\protect\cite{Chiari90} and measured
low temperature ($T=4$K) lattice constants. The lattices of
interacting spins formed by the bonds are shown on
Fig.~\protect\ref{fig:lattices} and Fig.~\ref{fig:together}. The last
column shows the contribution of each spin pair to the ground state
energy. Bonds with the same numerical index are closely related in
terms of bond vectors and chemical environments such that their
contributions cannot and need not be distinguished in this
experiment. }
\label{tab:bonds}

\end{table}

\begin{figure}
    \centering\includegraphics[scale=1.0]{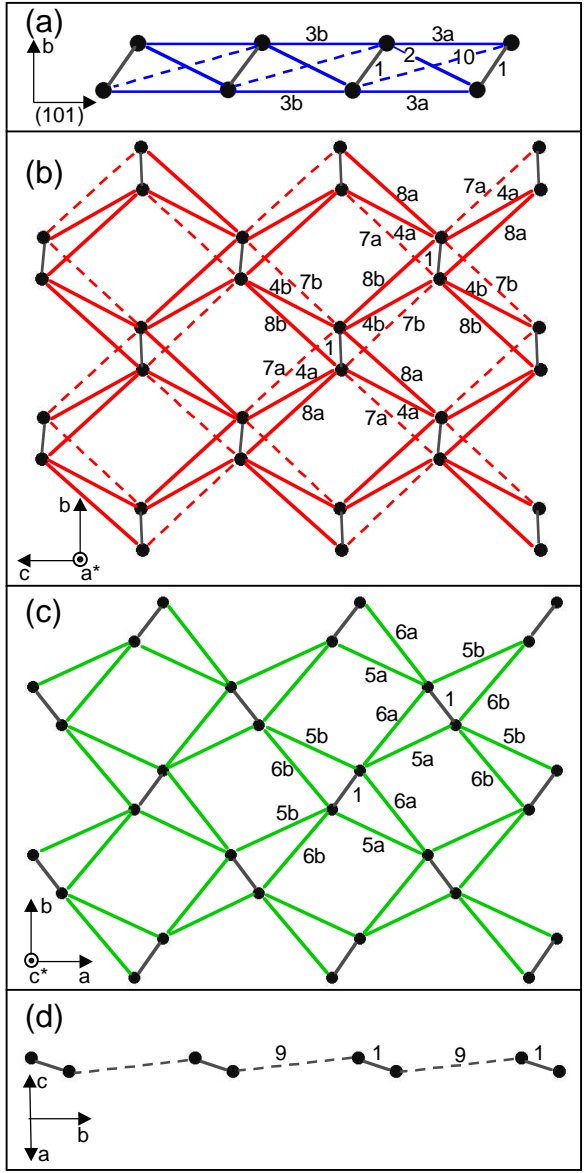}
\caption{\label{fig:lattices}
The four different types of lattices generated by the inter-molecular
interactions shown of Figs.~\protect\ref{fig:twomolecules}(a-d). The
molecules in Figs.~\protect\ref{fig:twomolecules} are oriented like
the spin pairs indicated by 1 in this figure. Dashed lines indicate
interactions that are expected to be very weak. Bond numbers refer to
Table 1. Sub-lattice color coding coincides with that of
Fig.~\protect\ref{fig:together}.
}
\end{figure}

The intra-molecular exchange interactions in CuHpCl do not create an
extended lattice. Because the molecular units in CuHpCl interact
solely through hydrogen bonding, the dimensionality and the overall
nature of magnetic interactions in this system are entirely determined
by hydrogen bonding mediated exchange interactions. Owing to the
slight deviation of the CuHpCl molecules from centro-symmetry,
molecular pairs come in two flavors in the CuHpCl crystal structure,
which we denote by a and b. Neglecting flavor distinctions, there are
four configurations for molecular pairs in direct contact as shown in
Fig.~\ref{fig:twomolecules}. The sublattices generated by each of
these pairs are shown in Fig.~\ref{fig:lattices}. The numbers on
Fig.~\ref{fig:lattices} indicate possible spin exchange interactions
mediated by hydrogen bonding and additional information about each of
these is listed in Table 1. Each molecule is part of two molecular
pairs of the type shown in Fig.~\ref{fig:twomolecules}(a). The three
different exchange interactions associated with this molecular pair
are denoted 2, 3, and 10 in Fig.~\ref{fig:lattices} and Table 1.
Bonds 2 and 3 proceed through a Cu-N-H-Cl-Cu path with H-Cl distances
ranging from 2.3 \AA\ to 2.7 \AA . Bond 10 has a similar exchange
pathway as for bonds 2 and 3 but must in addition traverse the entire
Cu($\mu$-Cl)$_2$Cu complex. The corresponding exchange interaction
should therefore be significantly weaker than for bonds 2 and 3. Each
molecule is also part of four pairs of the type shown in
Fig.~\ref{fig:twomolecules}(b). Exchange pathways involving
Cu-N-C-H-Cl-Cu mediate two different exchange interactions that we
denote 4 and 7, with H-Cl bond distances ranging from 2.6 \AA\ to 2.8
\AA. Bond 7 should be negligible, however, as it involves an apical
Cu-Cl contact. The same molecular pairs also afford an exchange
interaction denoted by 8, which proceeds through a Cu-Cl-H-C-H-Cl-Cu
pathway with H-Cl bond distances in the range 2.6-3.0 \AA.  Each
molecule also is part of four molecular pairs of the type shown in
Fig.~\ref{fig:twomolecules}(c). The exchange pathway involves either
Cu-Cl-H-C-N-Cu or Cu-N-H-H-C-N-Cu with H-Cl and H-H bond lengths
ranging from 2.7 \AA\ to 3.1 \AA . The corresponding exchange
interactions are denoted 5 and 6. Finally, each molecule is also part
of two molecular pairs of the type shown in
Fig.~\ref{fig:twomolecules}(d) with displacement vectors $\pm \bf b$.
The exchange path passes through two 1,4 diazacycloheptane rings so
the corresponding interactions are likely to be negligible,
especially considering the lower coordination number for this
interaction.

The molecular pairs in Fig.~\ref{fig:twomolecules}(a) and (d) yield
one-dimensional lattices extending along the [101] and [010]
directions respectively.  Inter-molecular interactions corresponding
to Figs. 2(b) and (c) on the other hand yield $\bf b-c$ and $\bf a-b$
spin planes with surface normals $\bf a$* and $\bf c$* respectively.
Unfortunately it is difficult to predict the strength of hydrogen
mediated exchange interactions, which reportedly can range from 0.1
meV to greater than 10 meV depending on the chemical
environment.\cite{yamada} Previous papers on CuHpCl worked under the
assumption that the inter-molecular bonds depicted in
Fig.~\ref{fig:twomolecules}(a) are dominant, leading to the ladder
model shown in Fig.~\ref{fig:lattices}(a). While we cannot provide
firm quantitative information about the magnitude of inter-molecular
exchange interactions, we shall present evidence that the 8
inter-molecular pairs of the type shown in Figs 2(b)-(c) when
combined are energetically more significant than those shown in
Fig.~\ref{fig:twomolecules}(a), and that the intra-molecular spin
pair (1) is in a frustrated configuration.

\section{Experimental Techniques}

  The powder sample studied consisted of 5.87
grams of deuterated CuHpCl.  To produce this sample, perdeuterated
1,4-diazacycloheptane was first prepared following a previously
published
method.  \cite{ChemAbstr} The $d_{6}$-dibromopropane and
$N,N^{\prime}$-dibenzylethylene-$d_{4}$-diamine precursors required
for this synthesis were prepared from commercially available
1,3-$d_{6}$-propanediol and ethylene-$d_{4}$-diamine,
respectively.  The CuHpCl powder was obtained by rapid cooling from
$40^{\circ}$C to $0^{\circ}$C of 1:1 molar solutions of anhydrous
$\mathrm{CuCl_{2}}$ and the deuterated 1,4-diazacycloheptane dissolved
in the minimum amount of deuterated methanol.  Prompt gamma neutron
activation analysis performed on a portion of the sample confirmed
$95.0(1)\%$ deuteration.

The single crystal measurements were performed on a composite sample
with a total mass  $m \approx 110$ mg.  This sample consisted of four
hydrogenous single crystals, mutually aligned to within 4.5 degrees.
These crystals were obtained by diffusive growth from
$\mathrm{CuCl_{2}}$ and 1,4-diazacycloheptane in methanol.
\cite{Ohta99} We have produced crystals as large as 33 mg using this
method.

Inelastic neutron scattering measurements on both the powder and
single-crystal samples were performed using the SPINS cold neutron
triple axis spectrometer at the National Institute of Standards and
Technology in Gaithersburg, Maryland.  For the powder experiment, the
horizontal beam collimation before the sample was $50^{\prime}/k_{i}$
(\AA$^{-1}$) - 80$^{\prime}$.  Scattered neutrons in the energy range
$2.6$ meV $\le E_{f} \le 3.7$ meV were Bragg reflected by a flat, 23
cm wide by 15 cm high
pyrolytic graphite analyzer [PG(002)] at a distance of 91 cm from the
sample position.  The analyzer was followed by an $80^{\prime}$
radial collimator and a position-sensitive detector.  Cooled Be and
BeO
filters were employed before and after the sample, respectively.  Data
were collected by scanning the scattering angle $2 \theta$ in the
range $7^{\circ}$ to $114^{\circ}$ at fixed incident energy $E_{i}$.
Scans at $E_{i} =$ 4.0, 4.34, and 4.84 probed inelastic scattering for
energy transfer $0.4$ meV $ \le \hbar \omega \le 2.09$ meV and
momentum transfer 0.2 \AA$^{-1}$ $ \le Q \le 2.35$ \AA$^{-1}$, with
average full width at half maximum (FWHM) resolutions $\delta \hbar
\omega = 0.14$ meV and $\delta Q = 0.014$
\AA$^{-1}$.\cite{ChesserAxe} Backgrounds due to
incoherent scattering from the analyzer and low-angle air scattering
of
the incident neutron beam were measured separately.  After subtracting
these, the data were converted to the normalized
scattering intensity $\tilde{I}(Q, \hbar \omega)$ using the measured
incoherent elastic scattering from the sample following the procedure
detailed in Ref.~\cite{Hammar98}.

For the single-crystal measurements, the horizontal beam collimation
before the sample was $50^{\prime}/k_{i}$ (\AA$^{-1}$) -
80$^{\prime}$.  A liquid nitrogen cooled BeO filter was placed after
the sample and data were collected at fixed final energy $E_{f} = 3.7$
meV while scanning the incident energy in the range $3.95 \le E_{i}
\le 5.45$ meV. A horizontally-focusing pyrolytic graphite analyzer
with horizontal and vertical acceptance angles of 5$^{o}$ and 7$^{o}$
respectively was
used in conjunction with a single cylindrical detector, which
subtended an angle of 4$^o$ in the horizontal plane to an area
element of the analyzer.   In this
configuration the average
instrumental energy resolution for the energy transfer range of 0.75
to 1.25 meV was $\delta \hbar \omega \approx 0.17(1)$ meV.
Representative values of the projected FWHM wavevector resolution
for the constant-Q scans performed throughout the measurement are
$\delta Q_{\parallel}=0.081$ \AA$^{-1}$ and $\delta
Q_{\perp}=0.065$ \AA$^{-1}$ for the components of the wavevector
transfer along the principal directions of the resolution ellipse at
${\bf Q} = (100)$ and
$\hbar \omega = 1$ meV.
Scans at constant wavevector transfer were performed
in both the ($h0l$) and ($hk0$) reciprocal lattice planes.

\section{Results}
\subsection{Powder Sample Measurements}

\begin{figure}
    \centering\includegraphics[scale=0.605]{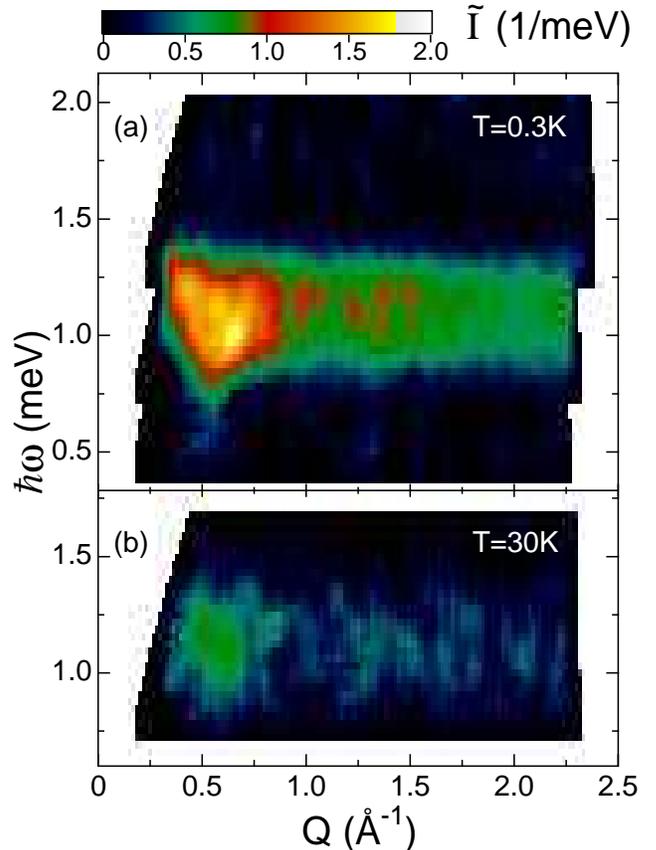}
\caption{\label{fig:colorplot}Powder inelastic
neutron scattering intensity $\tilde{I}(Q,\hbar\omega)$ for CuHpCl at
(a) T=0.3
K, and  (b) T=30 K obtained by combining measurements at $E_{i} =$
4.84, 4.3, and 4.0 meV.
The figure was produced by binning the data in
bins of size
$\delta \hbar\omega=0.03 $ meV and $ \delta Q=0.016$ \AA $^{-1}$
and then coarse-grain averaging to set the effective resolution to
$\Delta \hbar\omega=0.2 $ meV and $ \delta Q=0.08$ \AA $^{-1}$
}
\end{figure}

Figure \ref{fig:colorplot} shows the  normalized scattering intensity
$\tilde{I}(Q,\hbar \omega)$ for CuHpCl at
$T = 0.3 $ K and $T=30$ K.
At $T=0.3$ K, there is a band of inelastic scattering
in the range of energy transfer $0.9$ meV $ < \hbar \omega <$ 1.4 meV,
consistent with our previous measurements on a hydrogenous powder
sample.
\cite{Hammar98} At $T = 30$ K, the
intensity in this Q-$\hbar\omega$ range is diminished, which confirms
that the corresponding inelastic scattering cross section is
magnetic.

\begin{figure}
    \centering\includegraphics[scale=0.75]{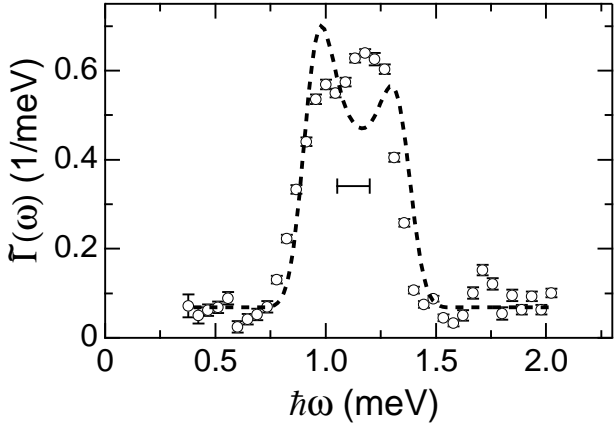}
\caption{\label{fig:intvse}Wavevector-averaged
scattering intensity vs. energy
transfer for CuHpCl.  The region of integration is limited to
 0.3 \AA$^{-1}$ $< Q < $ 2.3 \AA$^{-1}$. The horizontal bar indicates
 the FWHM energy resolution. The dashed line shows the resolution
convoluted spectrum for the previously accepted spin ladder
model.\protect\cite{Hammar98}}
\end{figure}

Figure \ref{fig:intvse} shows the wavevector averaged
scattering intensity
\begin{equation}
{\tilde I}(\omega)=\frac{\int Q^2dQ{\tilde{I}}(Q,\omega)}{\int Q^2dQ}
.
\end{equation}
This data is a measure of the magnetic density of states.
One-dimensional magnets have pronounced van Hove singularities at the
upper and lower bounds of the magnetic excitation spectrum. When
convolved with the energy resolution function, such singularities
would produce the spectrum shown by the dashed line in
Fig.~\ref{fig:intvse}.\cite{Hammar98}
The inconsistency between model and data provides a first indication
that CuHpCl is not a one dimensional spin system. A previous
experiment led to the same conclusion\cite{Hammar98} and as it was
done on a hydrogenous sample while the present sample is
deutererated, the comparison of these experiments indicates that
deuteration does not significantly alter magnetism in CuHpCl.
Fig.~\ref{fig:colorplot}(a) also shows that inelastic magnetic
neutron scattering from CuHpCl is strongly peaked
near $Q_0 \approx 0.6$ \AA$^{-1}$, with
a secondary maximum near $1.3$ \AA$^{-1}$. The powder averaged
scattering intensity from a single spin dimer with spacing $d$ has a
peak for $Q_0d=1.43\pi$.\cite{furrer} The data thus indicate singlet
formation in CuHpCl between spins separated by $d\approx
1.43\pi/Q_0=7.5$ \AA. This result is also inconsistent with the
ladder model for CuHpCl, as the dominant bond in this model is the
intra molecular Cu pair (Fig.~\ref{fig:onemolecule}) whose spacing is
only $d_1=3.376$ \AA .

\subsection{Single Crystal Measurements}

\begin{figure}
    \centering\includegraphics[scale=0.9]{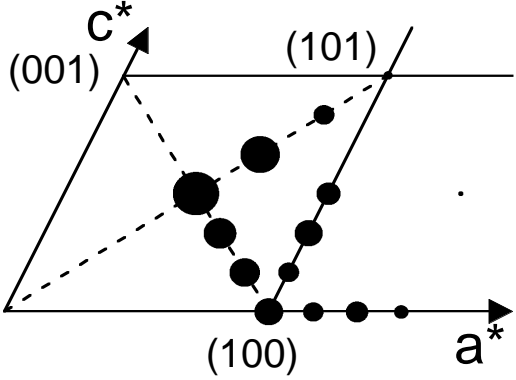}
\caption{\label{fig:qspacemap} Locations in the $(h0l)$ plane
at which single-crystal inelastic neutron
scattering data was obtained for CuHpCl.  The diameters of the points
are proportional to the measured first moment of the data, and show
that the dominant satisfied magnetic bond is parallel to [101].}
\end{figure}

\begin{figure}
    \centering\includegraphics[scale=0.55]{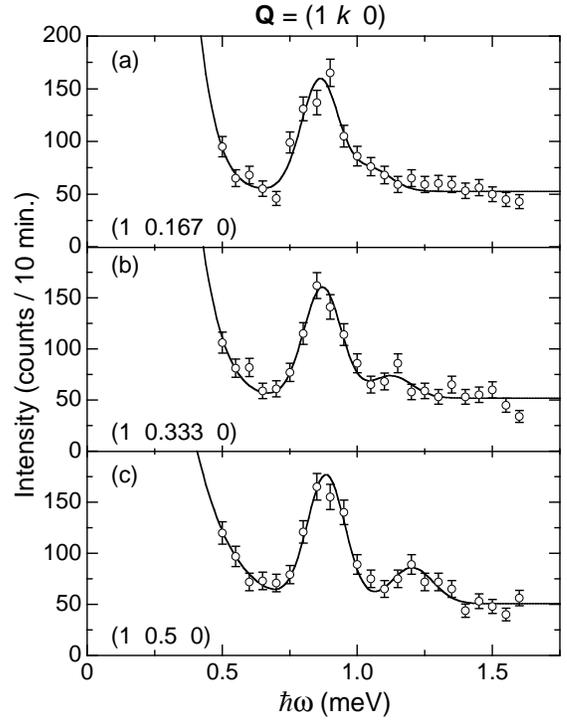}
\caption{\label{fig:onek0scans}Inelastic neutron scattering data for
CuHpCl indicating less than 0.05 meV dispersion along the $(1k0)$
direction.
The scan at $(1,0.5,0)$ shows the presence of two modes in
the excitation spectrum.  Solid lines
are fits to resolution limited gaussians.}
\end{figure}

\begin{figure}
    \centering\includegraphics[scale=0.55]{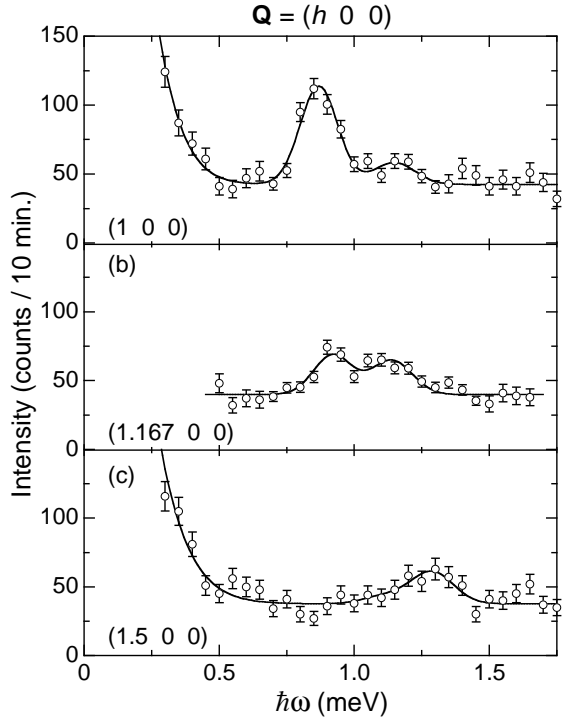}
\caption{\label{fig:h00scans}Inelastic neutron scattering data for
CuHpCl showing dispersion along $(h00)$, and perpendicular to the
lattice in Fig.~\protect\ref{fig:lattices}(b).  Solid lines
are fits to resolution limited gaussians.}
\end{figure}

\begin{figure}
    \centering\includegraphics[scale=0.55]{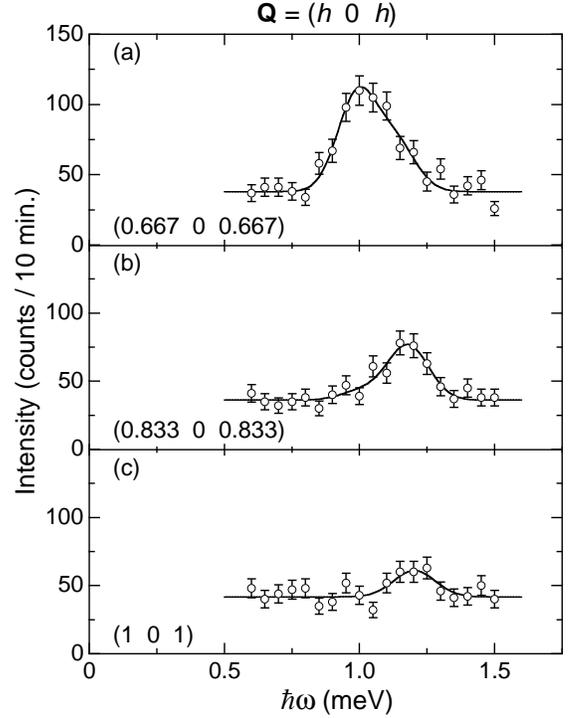}
\caption{\label{fig:h0hscans}Inelastic neutron scattering data for
CuHpCl indicating dispersion along the ($h0h$) direction.  Solid lines
are fits to resolution limited gaussians.}
\end{figure}

\begin{figure}
    \centering\includegraphics[scale=0.55]{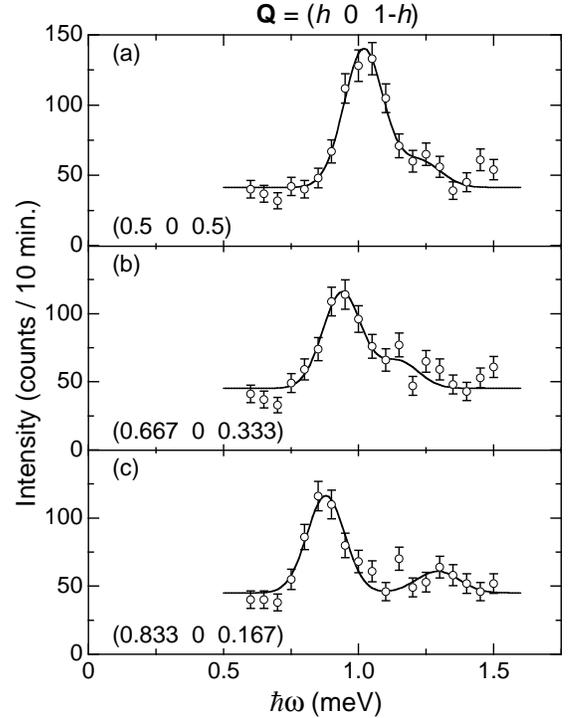}
\caption{\label{fig:h01-hscans}Inelastic neutron scattering data for
CuHpCl showing dispersion along the ($h,0,1-h$) direction,
which is perpendicular the spin ladder in
Fig.~\protect\ref{fig:lattices}(a).  Solid lines
are fits to resolution limited gaussians.}
\end{figure}

\begin{figure}
    \centering\includegraphics[scale=0.55]{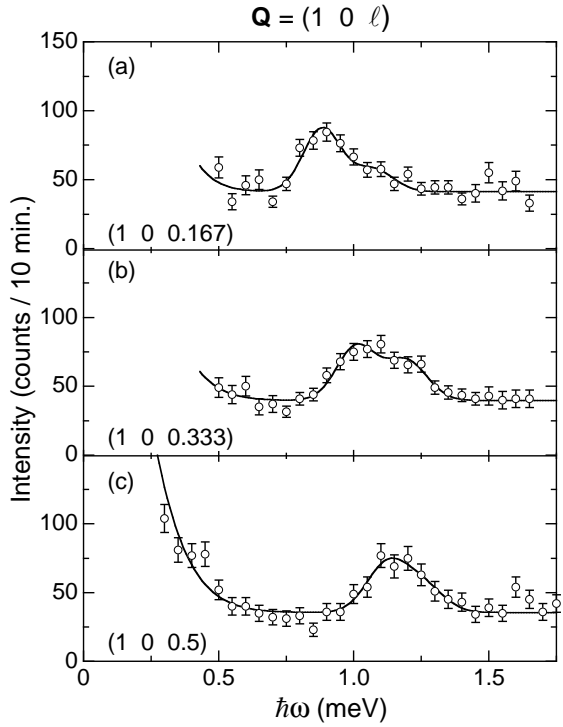}
\caption{\label{fig:10Lscans}Inelastic neutron scattering data for
CuHpCl. (a)-(c) show dispersion along the $(10l)$ direction and
perpendicular to the lattice in Fig.~\protect\ref{fig:lattices} (c).
Solid lines
are fits to resolution limited gaussians.}
\end{figure}

\begin{figure}
    \centering\includegraphics[scale=0.6]{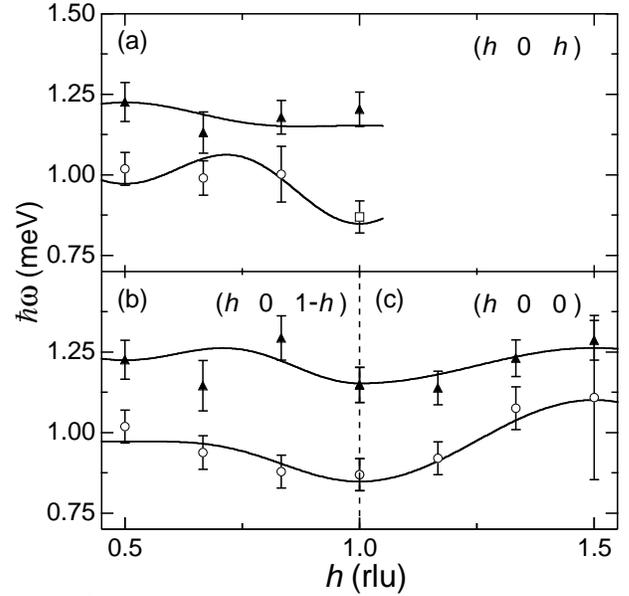}
\caption{\label{fig:disph}Dispersion of magnetic excitations in CuHpCl
derived from the fits shown in Figs.~\protect\ref{fig:h00scans},
\protect\ref{fig:h0hscans}, and \protect\ref{fig:h01-hscans}. Open
circles represent the lower energy mode, solid triangles represent
the higher energy mode.
Solid lines are phenomenological fits to a dispersion relation
satisfying Bloch's theorem: $\hbar \omega ({\bf Q}) =  A_{0} +
A_{1}\cos{2 \pi h}+ A_{2}\cos{2 \pi l}
+ A_{3}\cos{2 \pi (h+l)}+ $ $A_{4}\cos{2 \pi (h-l)}$. Assuming no mode
crossing, the constants are 1.00(2) meV, -0.04(2) meV, -0.02(2) meV,
-0.07(2) meV, and -0.02(2) meV for the lower mode and 1.21(2) meV,
-0.03(2) meV, 0.00(2) meV, 0.01(2) meV, and -0.03(2) meV for the
upper mode respectively.}
\end{figure}

\begin{figure}
    \centering\includegraphics[scale=0.6]{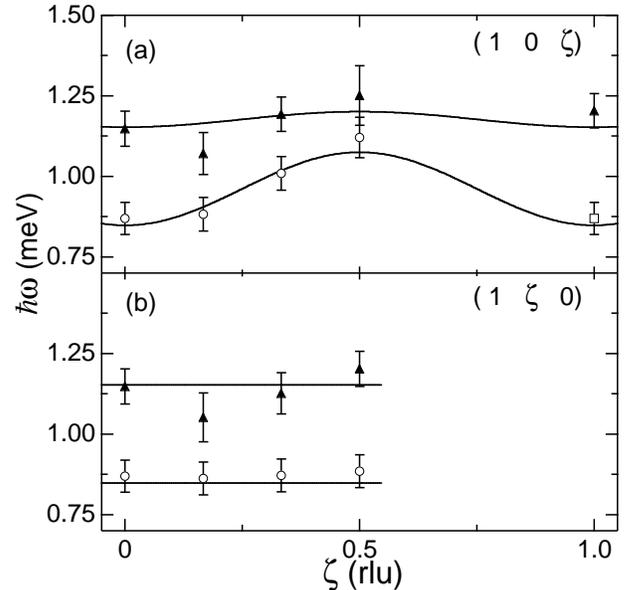}
\caption{\label{fig:dispkl}Dispersion of magnetic excitations in
CuHpCl derived from the data shown in Figs.~\ref{fig:onek0scans} and
\ref{fig:10Lscans}.
Open circles represent the lower energy mode, solid triangles
represent the higher energy mode, and the open square point at (101)
represents the energy of the dominant mode at (100)
 translated by ${\bf \tau} = (001)$.   Solid
lines are phenomenological fits as described in the caption to
Fig~\ref{fig:disph}.}
\end{figure}

Inelastic scattering measurements were carried out for wavevector
transfers at the locations in the
$(h0l)$ plane indicated on Fig.~\ref{fig:qspacemap}, as
well as
along the line $(1k0)$ perpendicular to this plane.  Figures
\ref{fig:onek0scans}-\ref{fig:10Lscans} show the data so
obtained, while Figs. \ref{fig:disph} and \ref{fig:dispkl} summarize
the corresponding dispersion relations derived by fitting the
constant wavevector scans to resolution limited peaks.

When dynamic correlations are dominated by a single dimerized
spin pair, there is a well tested RPA theory that can account for many
aspects of the magnetic excitation spectrum.
\cite{Leuenberger84,Sagaso97,Cavadini99,Cavadini00} As we shall show
in the following, each spin in CuHpCl takes part in several strongly
correlated spin pairs, so the RPA theory is not applicable in its
present form.
However, inspection of the excitation spectra and the crystal
structure leads to important insights concerning the magnetism in
CuHpCl.

We first note that there are two modes in the
magnetic excitation spectrum.  This is most easily seen at ${\bf
Q} = (1,0.5,0)$ in Fig.~\ref{fig:onek0scans}(c), where two
resolution-limited peaks are clearly visible at $\hbar \omega = 0.88$
and 1.2 meV. Two modes are also visible at other wavevectors such
as ${\bf Q} = (1.167,0,0)$ in Fig.~\ref{fig:h00scans}(b).  In
addition, Fig.~\ref{fig:h0hscans}(a) and Figs.
\ref{fig:10Lscans}(a-c) show broad or asymmetric peaks that are well
described by a superposition of two resolution-limited Gaussian
peaks.

The spin ladder and alternating chain models for CuHpCl corresponding
to Fig.~\ref{fig:lattices}(a) and \ref{fig:lattices}(d), respectively,
both have a single degenerate triplet excitation and are therefore
inconsistent with the observed two modes. This is true despite the
fact that Cu($\mu$-Cl)$_2$Cu complexes in CuHpCl come with two
different orientations as the inter-molecular interactions in
Fig.~\ref{fig:twomolecules}(a) and \ref{fig:twomolecules}(d) only
couple molecules with like orientations. However, the inter-molecular
interactions in Fig.~\ref{fig:twomolecules}(b) and
\ref{fig:twomolecules}(c) link molecules with different orientations
to form lattices with two molecules per unit cell
[Figs.~\ref{fig:lattices}(b) and \ref{fig:lattices}(c)]. If the
ground and excited states maintain the full symmetry of the
paramagnetic molecule and the spectrum is dominated by resonant
modes, then any lattice except those in Figs. 3(a) and (d) has two
triplet excited states consistent with the experimental data. This
means that the lattices shown in Fig.~\ref{fig:lattices}(b) and/or
\ref{fig:lattices}(c) are essential parts of the interacting spin
system in CuHpCl.

This conclusion is reinforced by Figs.~\ref{fig:h01-hscans} and
\ref{fig:disph}(b), which show dispersion along the $(h 0 1-h)$
direction [equivalent to $(h0\bar{h})$]. With displacement vectors
along [101] and [010], the molecular pairs in
Fig.~\ref{fig:twomolecules}(a) and \ref{fig:twomolecules}(d) cannot
yield dispersion along this direction in reciprocal space. The
implication is again that the inter-molecular interactions shown in
Figs \ref{fig:twomolecules}(b) and/or \ref{fig:twomolecules}(c), and
hence the lattices shown in Figs \ref{fig:lattices}(b) and/or
\ref{fig:lattices}(c), are essential parts of the cooperative magnetic
network in CuHpCl.
Figures \ref{fig:h00scans} and \ref{fig:disph}(c) show that there is
also dispersion along the $\bf a$* direction. This implies that the
interactions corresponding to Fig.~\ref{fig:twomolecules}(b) and
\ref{fig:lattices}(b) cannot be the only relevant inter-molecular
interactions in the problem.
Figures \ref{fig:10Lscans} and \ref{fig:dispkl}(a) show that there is
dispersion along the $\bf c$* direction. Consequently the
interactions corresponding to Fig.~\ref{fig:twomolecules}(c) and
\ref{fig:lattices}(c) also cannot be the only relevant inter-molecular
interactions. By inference, at least two of the four
types of inter-molecular interactions in Fig.~\ref{fig:twomolecules}
are important to cooperative magnetism in CuHpCl, and the
interactions in Figs.~\ref{fig:twomolecules}(b) and/or
\ref{fig:twomolecules}(c) must be part of the group.

Figure  \ref{fig:together} shows a projection of the lattice of
interactions in CuHpCl along the $\bf b$ direction. The color-coding
is consistent with Fig.~\ref{fig:lattices} indicating bonds
associated with different inter-molecular interactions. As the
lattices of Fig.~\ref{fig:lattices} interpenetrate,
a corollary to the above is that CuHpCl is a three-dimensional
interacting spin system. Nonetheless Figs. \ref{fig:onek0scans} and
\ref{fig:dispkl}
show that there is less than 0.05 meV dispersion
along ${\bf Q} =(1 k 0)$. Based on the discussion above, the absence
of dispersion along $\bf b$ cannot be due to lack of interactions
that couple the system magnetically. A likely alternate explanation
is that geometrical frustration localizes the excitations as has been
observed in other geometrically frustrated
systems.\cite{zncro,kageyama00,totsuka}

\section{Analysis}

\subsection{Wavevector Dependent Intensity}

The dynamic spin correlation function ${\cal S}({\bf Q},\omega)$
obeys sum rules that can be used to draw additional model independent
conclusions about magnetism in CuHpCl. The total moment sum
rule\cite{love} provides an important check on whether the measured
scattering intensity accounts for all spins in the sample:
\begin{equation}
\frac{\int d^3{\bf Q} \int \hbar d\omega
\sum_{\alpha}{\cal S}^{\alpha,\alpha}({\bf Q},\omega )}{\int d^3{\bf
Q}}
= S(S+1).
\label{totalmoment}
\end{equation}
The first moment sum rule\cite{hohenberg74} provides a direct link
between raw data and interaction terms in the spin Hamiltonian.
\begin{eqnarray}
\hbar \langle \omega\rangle_{\bf
Q}&\equiv&\hbar^2\int_{-\infty}^{\infty}\omega  {\cal S}^{\alpha
\alpha}({\bf
Q},\omega
)d\omega \\
&=& -\frac{1}{3}\frac{1}{N}
\sum_{\bf r,d} J_{\bf d}\langle{\bf S}_{\bf r}\cdot{\bf S}_{\bf
r+d}\rangle
(1-\cos {\bf Q\cdot d})
\label{eq:firstmomsumr}
\end{eqnarray}
Here \{{\bf d}\} is the set of all bond vectors connecting a spin to
its neighbors, the index $\{{\bf r}\}$  runs over all $N$ spins.
The Hamiltonian is assumed to take the form
\begin{equation}
{\cal H}= \frac{1}{2}\sum_{\bf r,d} J_{\bf d}{\bf S}_{\bf r}\cdot
{\bf S}_{\bf r+d} ,
\label{eq:spinham}
\end{equation}

For a powder sample, the magnetic component of $\tilde{I}(Q, \hbar
\omega)$
is related to the spherically averaged dynamic spin correlation
function ${\cal S}(Q,\omega)$ by \cite{Hammar98}

\begin{eqnarray}
\tilde{I}_m(Q,\hbar\omega ) &=& 2 \int dQ^{\prime} \hbar d\omega
^{\prime}
{\cal R}_{Q\omega }(Q-Q^{\prime},\omega-\omega^{\prime}) \nonumber \\
& & |\frac{g}{2}f(Q^{\prime})|^2{\cal
S}(Q^{\prime},\omega^{\prime} ) ,
\label{eq:ItildeQw}
\end{eqnarray}

where ${\cal R}_{ Q \omega }(\delta Q, \delta \omega)$ is the
normalized instrumental resolution
function, \cite{ChesserAxe} $f(Q)$ is the magnetic form
factor\cite{formfac} for
Cu$^{2+}$, and $g$ is the average $g$-factor, which is $g = 2.083$ for
CuHpCl. \cite{ChabouPRB97} Carrying out the integration of
Eq.~\ref{totalmoment} for the
band of intensity in Fig.~\ref{fig:colorplot}(a)
yields  $0.7(1)$. The proximity of the value to $S(S+1)=3/4$
indicates that this band
accounts for most of the magnetic scattering from CuHpCl.
For a powder sample, the first moment $\hbar \langle
\tilde{\omega}\rangle_Q$ of
the measured quantity $\tilde{I}_m(Q,\hbar\omega )$ is related to the
spherical average of Eq.~\ref{eq:firstmomsumr}, and is given by
\begin{eqnarray}
\hbar \langle \tilde{\omega}\rangle_Q&\equiv&
\hbar^2\int_{-\infty}^{\infty}\omega  \tilde{I}_m(Q,\hbar\omega )
d\omega \nonumber \\
&=& -\frac{2}{3} |\frac{g}{2}f(Q)|^2
\frac{1}{N}\sum_{\bf r,d}J_{\bf d}\langle{\bf S}_{\bf r}\cdot{\bf
S}_{\bf r+d}\rangle
\left ( 1-\frac{\sin {Qd}}{Qd}\right ) ,
\label{eq:firstmomtilde}
\end{eqnarray}
where $d = |{\bf d}|$.
The first moment $\hbar \langle \tilde{\omega}\rangle_Q$ computed from
the data in Fig.~\ref{fig:colorplot}(a) is shown in
Fig.~\ref{fig:firstmoment}.  In  strongly dimerized systems, the term
arising from the intra-dimer bond dominates
the first moment.\cite{Xu00} In the strongly dimerized two-leg ladder
model
of CuHpCl, these are the intra-molecular rung bonds labeled by 1 in
Table 1 and depicted in Fig.~\ref{fig:lattices}(a).\cite{ChabouPRB97}
As shown by the dashed line in
Fig.~\ref{fig:firstmoment}, the Q-dependence arising from inserting
$d_{1}$ in Eq.~\ref{eq:firstmomtilde} is manifestly inconsistent with
the data.
Specifically, the $d_1=3.376$ \AA\ bond yields a maximum at higher Q
than
is observed in the experiment.  Thus, a longer bond (or bonds) must
give the dominant contribution to $\hbar \langle
\tilde{\omega}\rangle_Q$.
Fitting the data in Fig.~\ref{fig:firstmoment} to
Eq.~\ref{eq:firstmomtilde} with a single, variable bond length
$d_{\alpha}$
yields $d_{\alpha} = 7.5(2) $ \AA, and the dashed-dotted line shown
in Fig.~\ref{fig:firstmoment}. However, this model is still
inconsistent with the data. The salient discrepancy is that the ratio
of the low-Q peak intensity to that of the high Q plateau is about
twice larger in the data than in the model.
Such a large ratio can only be achieved by combining terms of varying
sign in Eq.~\ref{eq:firstmomtilde}. With appropriate spin spacings,
$d$, such terms can interfere destructively in the high Q plateau
while the low-Q regime is dominated by contributions from long bonds.
Indeed, the solid line in Fig.~\ref{fig:firstmoment}, which we shall
describe in greater detail below, corresponds to a model with both
positive and negative terms in Eq.~\ref{eq:firstmomtilde}. According
to this equation, the magnitude of the high-Q plateau is directly
proportional to the shift of the ground state energy below zero while
the peak height measures the strength of individual spin pair
contributions to the ground state energy. Hence, there is a direct
link between a large peak to plateau ratio in first moment data, and
frustrated interactions that raise the ground state energy.

\begin{figure}
    \centering\includegraphics[scale=0.5]{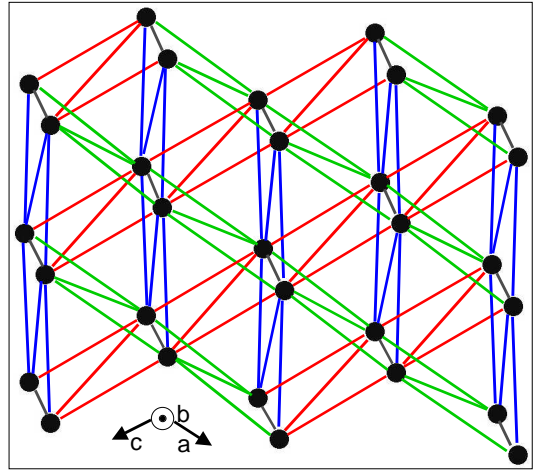}
\caption{\label{fig:together}  The combined lattice of spin-spin
interactions in CuHpCl viewed as a projection on the $\bf a-c$ plane
with the same color coding as for the individual sub-lattices in
Fig.~\ref{fig:lattices}.}
\end{figure}

The single crystal data help to distinguish
between the eight distinct spin pairs with Cu-Cu spacings in the
range 7 to 7.6 \AA. The first moment $\hbar
\langle \tilde{\omega}\rangle_{\bf Q}$ of the magnetic scattering
intensity,
$\tilde{I}_m({\bf Q},\hbar\omega )$, from a single crystalline sample
is given by
\begin{eqnarray}
\hbar \langle \tilde{\omega}\rangle_{\bf Q}&\equiv&
\hbar^2\int_{-\infty}^{\infty}\omega  \tilde{I}_m({\bf Q},\hbar\omega
)
d\omega \nonumber \\
&=& -\frac{2}{3} |\frac{g}{2}f(Q)|^2
\frac{1}{N}\sum_{\bf r,d}J_{\bf d}\langle{\bf S}_{\bf r}\cdot{\bf
S}_{\bf r+d}\rangle
\left ( 1-\cos{{\bf Q}\cdot {\bf d}}\right ) .
\label{eq:firstmomxtal}
\end{eqnarray}
Here we have neglected any spin space anisotropy, a reasonable
assumption for a spin-1/2 quantum spin liquid.
The first moment may be determined from gaussian fits to individual
spectra as follows
\begin{equation}
\hbar \langle \tilde{\omega}\rangle_{\bf Q} = \sum_{i}\tilde{I}({\bf
Q})_{i}
\hbar \omega_{i}  ({\bf Q}),
\label{eq:fmxtalmeas}
\end{equation}
where $\tilde{I}_{i}({\bf Q})$ is the integrated intensity for mode
$i$ at wavevector
${\bf Q}$ and $\hbar\omega_{i}({\bf Q})$ is the corresponding mode
energy. The wavevector dependence of $\hbar \langle
\tilde{\omega}\rangle_{\bf
Q}$ is illustrated in Fig.~\ref{fig:qspacemap} where the diameter
of the circles is proportional to $\hbar \langle
\tilde{\omega}\rangle_{\bf Q}$ and as conventional plots along
symmetry directions in  Figs.~\ref{fig:fmh} and \ref{fig:fmkl}.
There is an undetermined overall scale factor as the single crystal
data was not normalized in this experiment.
The first moment is seen to be
largest for $h\approx 0.5$ along the line $(h01-h)$.  Since the
magnitude of {\bf Q} in that part of reciprocal space is close to the
value 0.55 \AA$^{-1}$ where the peak in the first moment of the
powder data occurs, the powder and single crystal data are
consistent.  The strongest modulation in
$\hbar \langle \tilde{\omega}\rangle_{\bf Q}$ occurs along the $(h0h)$
direction, with weaker modulation also visible in other directions.
From the form of Eq.~\ref{eq:firstmomxtal}, this implies that the bond
vector that contributes most strongly to the first moment is
parallel to [101].

To determine the relative importance of the magnetic bonds, we
carried out a simultaneous
fit of the powder data to Eq.~\ref{eq:firstmomtilde} and the
single-crystal data to Eq.~\ref{eq:firstmomxtal} with a single set of
values for the bond energies, $J_{\bf d}\langle{\bf S}_0\cdot{\bf
S}_{\bf d}\rangle $. Because their contributions to the first moment
of the scattering data cannot be distinguished, we derive only an
average correlation term for spin pairs labeled with the same
numerical index in Table \ref{tab:bonds}. Such spin pairs would be
equivalent had the molecular unit possessed centro-symmetry. In
addition, there is close similarity between the chemical environments
along the exchange pathway of a and b labeled bonds. These facts lend
some support to the assumption that the corresponding bond energies
are similar.

\begin{figure}
    \centering\includegraphics[scale=0.7]{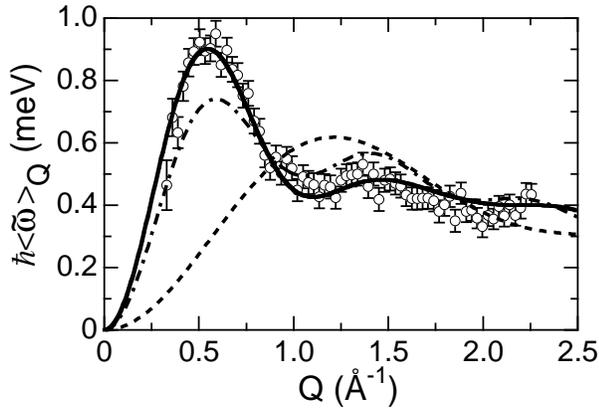}
\caption{\label{fig:firstmoment}First $\hbar\omega$ moment of
the inelastic powder data in
Fig.~\protect\ref{fig:colorplot}
versus wavevector transfer (Eq.~\ref{eq:firstmomtilde}).  The region
of integration is limited to the band of magnetic excitations from 0.7
to 1.5 meV.  The dashed-dotted line is a fit with a single dimer bond
length of 7.5(2) \AA.  The dashed line is a fit fixing the dimer
bond length to 3.376 \AA, corresponding to the intra molecular
Cu-spacing.  The
solid line is the best combined fit to the powder and single crystal
data.  Fit results are described in the text and enumerated in
Table~\ref{tab:bonds}}
\end{figure}

The results of the fit are given in the last column of
Table \ref{tab:bonds} and as solid lines in
Figs.~\ref{fig:firstmoment}- \ref{fig:fmkl}. It is important to note
the direct contribution of a spin pair to the ground state energy is
small when the exchange constant and/or the spin correlation function
is small. Furthermore, according to Eq.~\ref{eq:spinham}, negative
bond energies lower the ground state energy while positive terms
indicate a frustrated spin pair that raises the ground state energy.

\section{Discussion}
The most remarkable result from Table 1, is that the intra-molecular
bond 1 is in a frustrated configuration. The presence of a frustrated
spin pair with a short bond vector was anticipated based on
inspection of the raw powder first moment data. It is easily verified
that in a closed loop of interacting spins with an odd number of
antiferromagnetic exchange interactions, only an even number of spin
pairs can be satisfied. Bond 1 is part of no fewer than 5 near
neighbor bond triangles that are frustrated if all interactions are
antiferromagnetic. Figure~\ref{fig:lattices}(a) shows two of these
triangles: (1,3,2) and (1,3,10). Fig.~\ref{fig:lattices}(b) shows
bond triangles (1,4,8) and (1,4,7), while Fig.~\ref{fig:lattices}(c)
shows bond triangle (1,5,6). Of these, Table 1 clearly indicates that
the triangles in Fig.~\ref{fig:lattices}(b) are frustrated. Our
analysis indicates that bonds 4 and 8 are satisfied at the apparent
expense of bond 1. The energy derived for bond 7 while negative, is
not statistically significant. This is consistent with the
expectation from section~\ref{structure} that this is a weak exchange
interaction.

\begin{figure}
    \centering\includegraphics[scale=0.6]{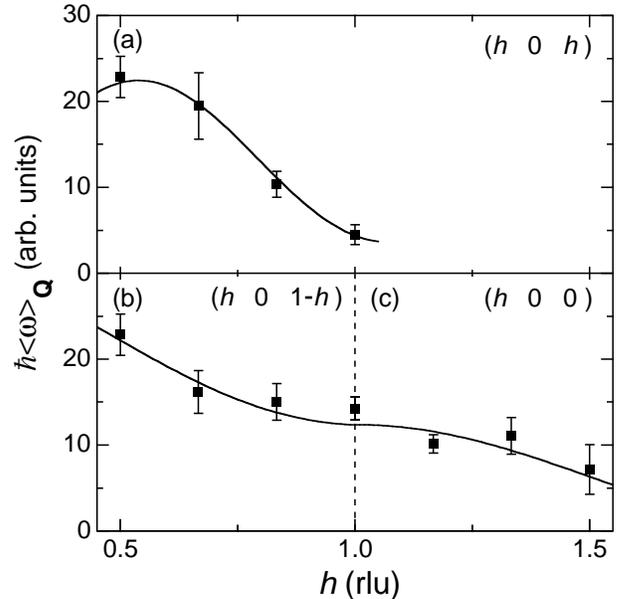}
\caption{\label{fig:fmh}First moment of the magnetic excitations in
CuHpCl derived from single-crystal data.  Curves are fits as described
in text.}
\end{figure}

\begin{figure}
    \centering\includegraphics[scale=0.6]{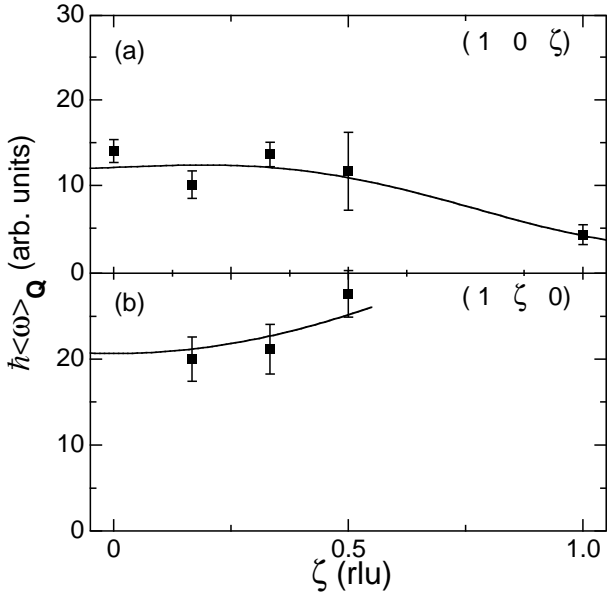}
\caption{\label{fig:fmkl}First moment of the magnetic excitations in
CuHpCl derived from single-crystal data.  Curves are fits as described
in text.  Data from the $(hk0)$ zone were scaled to data from the
$(h0l)$ zone using $(100)$ as a point of reference.}
\end{figure}

The connectivity of the lattice formed by bonds 1, 4, and 8 is that
of the geometrically frustrated Shastry-Sutherland model
(SSM)\cite{ssmodel} albeit with lower symmetry. Compared to the SSM,
which can be described as a square lattice with alternating diagonal
bonds on half the squares, the lattice shown by solid lines in
Fig.~\ref{fig:lattices}(b) corresponds to a tetragonal lattice formed
by bonds 4 and 8 with the ``diagonal" bonds 1 arranged as in the SSM.
When the square lattice exchange interactions in the SSM are less
than 70\% of the interactions across the diagonal, the ground state
of the SSM consists of singlets on every cross
bond.\cite{mullerhartmann} For stronger inter-dimer interactions,
there is a first order transition to square lattice N\'{e}el order.
Table 1 indicates that the SSM planes in CuHpCl have local spin
correlations resembling the N\'{e}el phase of the SSM.  The lattice
formed by bonds 1, 5, and 6 also has the topology of the SSM but
correlations in this plane do not readily map on a known phase of the
SSM.

It does seem surprising that spin pair 1 can provide a three times
larger positive contribution to the ground state energy than the
negative contributions from bonds 4 and 8. To determine whether this
is plausible, we examined a series of frustrated antiferromagnetic
spin-1/2 clusters. A central spin pair, $ {\bf S}_a, {\bf S}_b$ with
antiferromagnetic exchange constant $J>0$ is surrounded by $2n$
spins-1/2 ${\bf S}_1, {\bf S}_2, ... {\bf S}_{2n}$, which interact
with both pair members with equal antiferromagnetic exchange,
$J^\prime>0$. The spin Hamiltonian is given by
\begin{eqnarray}
{\cal H}&=&{\cal H}_0+{\cal H}_{2n}\\
{\cal H}_0&=&J{\bf S}_a\cdot{\bf S_b}\\
{\cal H}_{2n}&=&J^\prime\sum_{i=1}^{2n} {\bf S}_i\cdot ({\bf
S}_a+{\bf S}_b)
\end{eqnarray}
We examined clusters with $n$ ranging from 1 to 3. For each cluster
we diagonalized the spin Hamiltonian and determined the ground state
frustration index
\begin{equation}
f=-4n\frac{<0|{\cal H}_0|0>}{<0|{\cal H}_{2n}|0>}
\end{equation}
as a function of $x=J^\prime/J$ in the range where the central spin
pair is frustrated. We obtained maximum frustration indexes
$f_{max}=1,2,3$ for $n=1,2,3$ respectively.
The results show that a level of frustration similar to that observed
for the intra-molecular spin pair in CuHpCl is possible even for very
small spin clusters.

While the lattice of Fig.~\ref{fig:lattices}(b) provides an
explanation for the frustration of bond 1, it does not readily
account for the singlet ground state of CuHpCl. Local correlations in
these planes is N\'{e}el-like so the lattice left on its own might be
expected to have a gapless spectrum and long range order at low
temperatures. From Table 1 we see that bond 3, which generates the
lattice in Fig.~\ref{fig:lattices}(a), is in an un-frustrated
configuration. In fact, this bond provides the greatest contribution
towards lowering the ground state energy. Fig.~\ref{fig:together}
shows how the lattices of Figs. 3(a)-(c) intersect to form a three
dimensional lattice. We see that if the lattice in
Fig.~\ref{fig:lattices}(b) were in a N\'{e}el phase, then bond  3
only serves to strengthen such N\'{e}el order and extend it to three
dimensions. It is therefore difficult to see how the sub-lattices
from Figs.~\ref{fig:lattices}(a) and \ref{fig:lattices}(b) alone can
account for an isolated singlet ground state.

This leads us to the suggestion that bonds 5 and 6 could play a
significant role even if the corresponding bond energies appear to be
small. As is apparent from Fig.~\ref{fig:together}, bonds 5 and 6
close a molecular triangle that projects onto the $\bf a-c$ plane.
The description therefore emerges of a set of antiferromagnetic SSM
layers normal to $\bf a^*$ coupled in a frustrating triangular
lattice geometry to a second interpenetrating set of frustrated SSM
layers normal to $\bf b^*$.  A small alternation between the 3a and
3b bonds that couple layers could be an additional factor favoring
singlet formation on the stronger of these bonds.

The magnetic ground state energy may be calculated from
Eq.~\ref{eq:spinham} and the numbers in Table 1 to be -0.36 meV per
spin. For comparison, if all bonds were satisfied with the same
magnitude of spin correlations as observed the ground state energy
would be -1.17 meV per spin. The ratio between the actual ground
state energy and the latter upper bound on the energy in the absence
of frustration is 0.3. For comparison the ground state energy of the
SSM at the critical point separating the N\'{e}el phase and the dimer
phase is 0.26 times the energy that the spin system would have if all
bonds could simultaneously be engaged in singlet
formation.\cite{takushima} Hence, CuHpCl is at least as frustrated as
the SSM at its critical point.

\section{Conclusions}
In summary, we have presented inelastic neutron scattering data from
deuterated powder and hydrogenous single crystals of the
organometallic spin-1/2 magnet CuHpCl. Consideration of the
excitation spectra, the crystal structure, and the wave vector
dependence of the first moment leads to the conclusion that
spin-spin interactions in this system form a complex
three-dimensional lattice and not a spin ladder as previously
thought. While structurally one might expect CuHpCl to fall in the
spin-dimer class of quantum spin liquids, this appears not to be the
case. The spin gap in spin dimer systems is a consequence of the
singlet ground state of individual dimers. However, in CuHpCl the
intra molecular spin interaction is in a frustrated configuration so
this cannot be the dominant interaction in the problem. Furthermore,
if the intra-molecular interaction is antiferromagnetic, as expected,
then the ground state features triplet molecules that in and of
themselves cannot explain a gap in the excitation spectrum. We are
therefore led to the conclusion that the pronounced gap in the
excitation spectrum of CuHpCl is a consequence of the frustration
inherent to this particular three-dimensional network of
interactions. This is a surprising discovery as the symmetry of the
lattice is low. On the other hand, the structure clearly is riddled
with triangular units, and the connectivity between them is
relatively low. These ingredients are known to be important for
suppressing N\'{e}el order and promoting a spin liquid
state.\cite{Ramirez} Further progress in understanding the magnetism
of CuHpCl would benefit from accurate determination of H/D positions
using neutron scattering, followed by quantum chemical calculations
of exchange constants. More extensive measurements of the magnetic
excitation spectrum are also needed, but these must await progress in
crystal growth or neutron scattering instrumentation.

There are many organometallic quantum magnets that have been labeled
as quasi-one-dimensional, based largely on the observation of a spin
gap. Our experiments on CuHpCl have shown how neutron scattering from
single crystals as small as 0.1 g can be used to establish the
dimensionality and the basic nature of interacting spin systems. They
also show that insulating magnets with a gap in their excitation
spectrum may constitute a considerably more complex and diverse class
of interacting quantum many body systems than previously anticipated.

\section{Acknowledgments}
We thank R. Paul for help with neutron activation analysis.
This work was supported by  NSF Grants DMR-9801742 and DMR-0074571.
DHR
acknowledges the support of the David and Lucile Packard Foundation.
TL acknowledges the support of the Sloan and Dreyfus Foundations.
X-ray characterization was carried out using facilities maintained by
the JHU MRSEC under NSF Grant number DMR-0080031.  This work utilized
neutron research facilities supported by NIST and the NSF under
Agreement No. DMR-9986442.

\newpage


\end{document}